# Nonlinear Spectroscopic Study of Porphyrin Under cw and Femto-second Laser Pulse Excitation

A. Srinivasa Rao[1†], Alok Sharan[1*], N Venkatramaiah[2], and R Venkatesan[2]

[1]*Department of Physics, Pondicherry University, Puducherry, 605014, India*
[2]*Department of Chemistry, Pondicherry University, Puducherry, 605014, India*
[†]*Present address: Photonic Sciences Lab, Physical Research Laboratory, Navarangpura, Ahmedabad 380009, Gujarat, India*
[†]Email: *asvrao@prl.res.in*
[*]Email: *alok.phy@pondicuni.edu.in*

**Abstract:** Single Beam Transmittance (SBT) was used as nonlinear spectroscopic tool to investigate the absorption cross-sections and lifetimes of Tetra Phenyl porphyrin ($H_2TPP$) and its $OH^-$ group derivative ($H_2TPP(OH)_4$) doped in boric acid glass (BAG). We have used 671 nm wavelength as exciting wavelength for both CW (incident intensity up to $10^{10}$ W/cm$^2$) and femto-second laser pulse (up to fluence of $10^2$ mJ/cm$^2$). Under cw laser excitation, $H_2TPP$ doped BAG demonstrates Double Saturable Absorber (DSA) behavior whereas $H_2TPP(OH)_4$ doped BAG act as Revere Saturable Absorber (RSA). Rate equation model espouses to extract the spectroscopic parameters from the experimental data. Excited state life times and absorption cross-sections were obtained as parameters for theoretical fit on SBT data. Porphyrin molecules act as four level systems under cw laser excitation, whereas in the presence of femto-second laser excitation they act as a two level system. We have derived the equations for transmitted energy through the material in the presence of femto-second laser illumination. Both systems viz., $H_2TPP$ and $H_2TPP(OH)_4$ doped BAG behaves as saturable absorbers when excited by femto-second laser pulses.

## 1. Introduction

Laser spectroscopy has pushed the limit of resolution to unprecedented levels and thereby improved understanding of the dynamics of the population, under laser illumination. Molecular interaction with light is generally linear under excitations with low intensity of light. The nonlinear optical interaction due to intense light illumination can be described by simultaneous interaction of a molecule with more than one photon [1-8]. The spectroscopic parameters which are difficult to obtain through linear spectroscopic techniques can be measured by nonlinear spectroscopic techniques like saturation spectroscopy, pump-probe technique, wave mixing, single beam transmittance (SBT), [9-15] etc. These processes could involve single and/or multi-photon absorption.

In order to obtain spectroscopic parameters through SBT or z-scan experiment, first of all, we need to know how many parameters contribute to the nonlinear optical process at the intensity probed. Inputs from linear characterization data (like absorption spectra) are used to model the SBT data. It is easy then to estimate the other parameters especially excited states properties [16]. In this report, we present nonlinear optical study on tetra-phenyl porphyrin molecule and it derivative doped in BAG [17-23]. We have estimated the spectroscopic parameters from the SBT data at 671nm wavelength, which lies within the absorption band. The study has been carried out under cw and femto-second laser excitation.

## 2. Sample

Porphyrin and its derivative such as 5, 10, 15, 20-*meso*-tetrakis phenyl porphyrin (H$_2$TPP) and 5, 10, 15, 20-*meso*-tetrakis (4-hydroxyphenyl) porphyrin (H$_2$TPP(OH)$_4$) powders of appropriate concentration were mixed thoroughly with boric acid powder and heated in a crucible to 150$^0$C. Melt formed was transferred on the pre-heated glass slides and then sandwiched between glass slides and rapidly quenched [24, 25]. For comparative study, we have used same concentration 4×10$^{-5}$ M for both H$_2$TPP and H$_2$TPP(OH)$_4$ in boric acid. The concentration is carefully chosen so that the UV-Vis absorption spectra do not show any effect of agglomerations of dyes [25]. We have used BAG as a host dielectric matrix because of high bond strength, low cation strength and smaller heat required for fusion. Trapping of these organic dyes in the rigid glassy network increased the life-time of excited states as compared with their liquid form [26]. The nonlinear optical properties of the molecules hosted in such a rigid environment are probed by cw as well as femto-second laser pulses. Energy level diagram of porphyrin molecules are better appreciated in the Jablonski diagram, where the singlet and triplet manifolds are shown slightly displaced, since the radiative transition among the manifolds is spin-forbidden [27]. The allowed transitions from S$_j$→S$_i$ (*j=i*+1) and T$_j$→T$_i$ (*j=i*+1) gives the fluoresence emission and forbidden transitions from T$_j$→S$_i$ (*j=i*+1) gives the phosphorescence emission.

### 3. CW laser excitation

#### 3.1. Model

We consider a four level energy system viz S$_0$, S$_1$, T$_1$ and T$_2$. The reason is that as soon as molecule in S$_0$ is excited to S$_1$ state, non-radiative processes and intersystem crossing would aid in transfer of population from S$_1$ to T$_1$. There is no radiative transfer between S$_1$ and T$_1$. For cw illumination this model, helps to describe the optical nonlinearity in the transmission whereas higher level singlet states are considered in the nonlinear absorption under ultra-short pulse illumination.

As shown in the figure 1, under CW laser excitation the population transfer from singlet (S$_1$) to triplet (T$_1$) by inter-system crossing and non-radiative decay ($\tau_{21}$). The long phosphorescence life-time, allows the material to saturate at lower intensity. Our experimental results imply that both H$_2$TPP and H$_2$TPP(OH)$_4$ doped BAG act as the four level systems (2 singlet & 2 triplet). Depending upon the ratio of the ground to excited state absorption cross-section this kind of four level systems shows either saturable or reverse saturable absorption [10, 12, 28 & 29]. In case of saturable absorption M D Rayo et al. has observed DSA in polymer films containing bacteriorhodopsin [30]. To explain our results, we have considered the four level model [16] with notations as depicted in figure 1. The population dynamics in the system can be understood through the rate equation formalism. The number density of molecules in each state "*i*" is considered to be $N_i$. The total number density of molecules is given as $N=\Sigma N_i$ and the fractional number density of each state is denoted as $n_i=N_i/N$, with $\Sigma n_i=1$. The pumping rate from $i^{th}$ to $j^{th}$ level is $W_{ij}=(I/h\nu)\sigma_{ij}$ and the decay time from $j^{th}$ to $i^{th}$ level is $\tau_{ji}$.

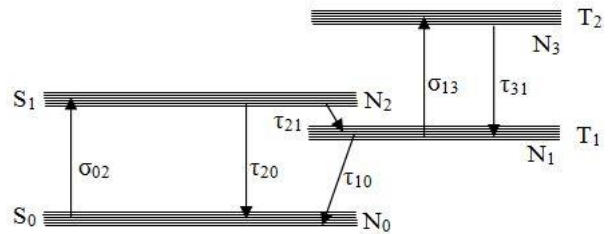

Figure 1 Energy levels of dye molecule in the presence of continuous wave illumination.

$$\frac{dn_0}{dt} = -W_{02}(n_0 - n_2) + \frac{n_1}{\tau_{10}} + \frac{n_2}{\tau_{20}} \quad (1a)$$

$$\frac{dn_1}{dt} = -W_{13}(n_1 - n_3) - \frac{n_1}{\tau_{10}} + \frac{n_2}{\tau_{21}} + \frac{n_3}{\tau_{31}} \quad (1b)$$

$$\frac{dn_2}{dt} = W_{02}(n_0 - n_2) - \frac{n_2}{\tau_{20}} - \frac{n_2}{\tau_{21}} \quad (1c)$$

$$\frac{dn_3}{dt} = W_{13}(n_1 - n_3) - \frac{n_3}{\tau_{31}} \quad (1d)$$

The transmittance of the material is given by $T = e^{-\alpha L}$, where $\alpha = N\Delta n_{02}\sigma_{02} + N\Delta n_{13}\sigma_{13}$; $\Delta n_{02} = n_0 - n_2$ and $\Delta n_{13} = n_1 - n_3$. The population in each level is obtained by solving rate equations under steady state condition. The media demonstrates saturation at two different values of intensity. This results in two "plateau" like function in the transmission curve at intensity corresponding to the values given by $I_{S01} = h\nu/\tau_{10}\sigma_{02}$ and $I_{S13} = h\nu/\tau_{31}\sigma_{13}$.

### 3.2. Experimental setup

Experimental setup used for SBT is shown in Figure 2. Laser source used is a cw diode laser (SDL-671-120T) emitting at 671nm wavelength with $TEM_{00}$ mode. To prevent the thermal effects affecting the nonlinear transmittance, we used optical chopper at 1 kHz chopping frequency. Two lens combinations are used to increase the intensity at sample position; where the first lens expands the beam to fill the second lens which then tight focuses the beam and thereby helps in achieving larger intensity. Sample was kept 5 mm away from the focus to avoid the laser induced damage. Neutral density filters are used as intensity attenuators to vary the incident intensity on the sample. The translation stage was fixed in such a way that data can be taken with and without sample by moving the sample perpendicular to beam propagation. Newport 918UV-OD3R photo diode with 842-PE optical power energy meter was used for optical measurements. First iris $A_1$ is used to avoid the instability in the diode laser by blocking reflections into it and second iris $A_2$ is used to prevent the back reflections in the nonlinear medium. Data were collected with integration time of four seconds. The fluctuations in the incident light are monitored with the reference diode ($PD_2$). The sample was found to have threshold damage intensity around $7.6 \times 10^6$ W/cm$^2$. figures 3 and 5 are plots of experimental results for $H_2TPP$ and $H_2TPP(OH)_4$ doped in BAG. These data were fitted with solutions of rate equations given in the equation 1 [16]. All the parameters used and obtained are presented in table 1. Populations for $H_2TPP$ and $H_2TPP(OH)_4$ doped BAG obtained by our theoretical model are shown in figures 4 and 6. We used the life-times and absorption cross-sections obtained from the theoretical fittings of experimental data presented in the figures 3 and 5 to study population distribution of $H_2TPP$ and $H_2TPP(OH)_4$ doped BAG (figures 4 and 6).

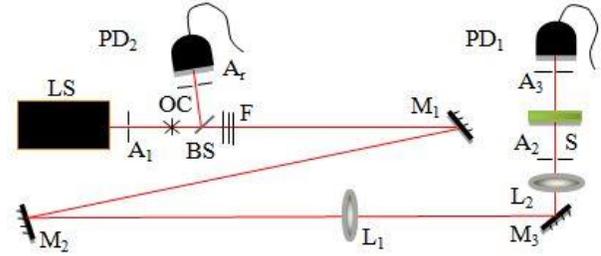

Figure 2 Experimental set up of SBT: LS → Laser source, OP → Optical chopper, F → Neutral density filters, $M_i$ → Mirror, , $L_i$ → Lens, S → Sample, PD → Photo diode, $A_i$ → Iris.

SBT profile of $H_2TPP$ (figure 3) doped in BAG shows saturable absorber behavior. This is due to large ground state absorption cross-section as compare with excited state absorption cross-section ($\sigma_{02} > \sigma_{13}$). Therefore second saturation occurs at higher intensity where complete bleaching occurs. Hence we say that it demonstartes saturable behavior twice and we term it as Double Saturation Absorrption (DSA).

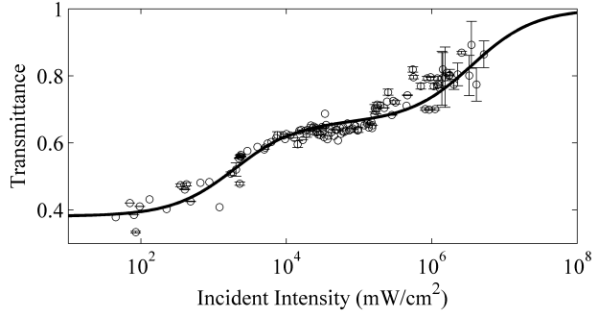

Figure 3 SBT profile of $H_2TPP$ ($4\times10^{-5}$ M) doped BAG where the circular experimental data is fitted with the theoretical curve.

In the transmittance profile (figure 4), we have plotted the fractional populations of different energy levels. Saturable absorption, $SA_1$ is formed due to trapping of molecules in the $T_1$ state ($n_1$) with depleted $S_0$ and $S_1$ ($n_0 \approx 0$, $n_2 \approx 0$) as $\tau_{20} > \tau_{21}$ and large $\tau_{10}$. As a result, the transmittance is constant in the $SA_1$ region. The $SA_2$ region is formed due to an equal number of molecules being shared by the $T_1$ and $T_2$ ($n_1 = n_3$) levels. With $n_0$ and $n_2$ population $\approx 0$, it achieves complete beaching leading to saturation of absorption viz., $SA_2$ region in the transmittance curve. Due to the availability of only low power cw laser with us, we were unable to reach the second saturation region. The transmittance profile for such system can be written as LA ($S_0$) → $SA_1$ ($T_1$) → $SA_2$ ($T_1$, $T_2$).

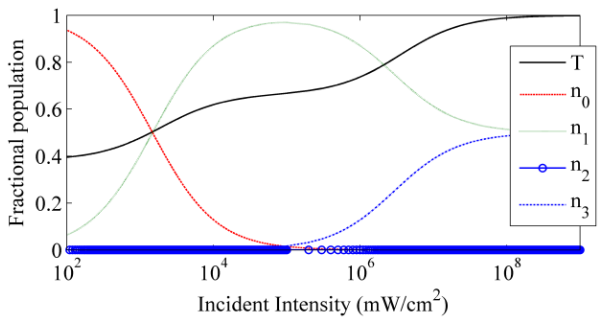

Figure 4. The population distribution of the porphyrin molecules among the energy levels in the transmittance profile of $H_2TPP$ ($4\times10^{-5}$ M) doped BAG.

In the presence of $OH^-$ group, the ground and excited states absorption cross-sections of porphyrin molecule are found to be larger as compared to free base porphyrin molecule. This increase is larger for the excited state absorption cross-section as compared with its ground state absorption cross-section. i.e., Presence of $OH^-$ group has led to alter the condition from $\sigma_{02} > \sigma_{13}$ to $\sigma_{02} < \sigma_{13}$. As a result, we observe the reverse saturable absorber (RSA) behavior in the $H_2TPP(OH)_4$ molecule doped in BAG [figure 5].

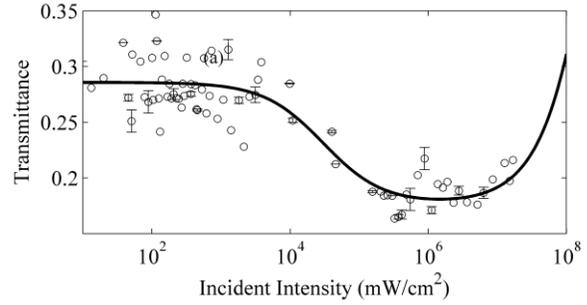

Figure 5 SBT of $H_2TPP(OH)_4$ ($4\times10^{-5}$ M) doped BAG where the circular experimental data is fitted with the theoretical curve.

In RSA region, molecules get trapped in the $T_1$ state ($n_1$) with depleted $S_0$ and $S_2$ ($n_0 \approx 0$, $n_2 \approx 0$) levels (figure 6). Under conditions $\sigma_{02} < \sigma_{13}$ and $\tau_{20} > \tau_{31}$ the absorption increases between $T_1$ and $T_2$ transitions. As a result, more energy is absorbed and it leads to RSA. As the incident intensity is further increased the system finally achieves saturation, when the population between the two triplet levels are equal leading to complete bleaching. The transmittance profile for this case is LA ($S_0$) → RSA ($T_1$, $T_2$) → SA ($T_1$, $T_2$).

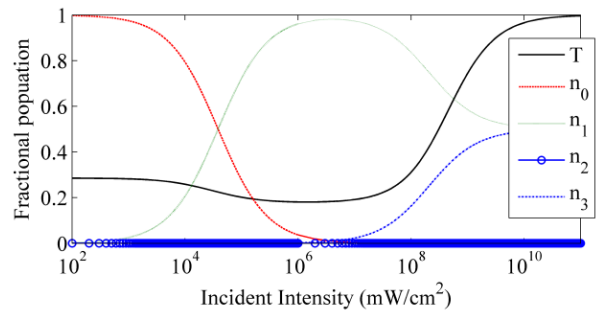

Figure 6 The population distribution of the porphyrin molecules in the transmittance profile of $H_2TPP(OH)_4$ ($4\times10^{-5}$ M) doped BAG.

The comparative results of H$_2$TPP and H$_2$TPP(OH)$_4$ doped BAG has been tabulated in table1. We have estimated the thickness of the films by DFWM experiment by monitoring the first minima. We have obtained the values of fluorescence life-time ($\tau_{20}$) experimentally, as both time $\tau_{20}$ and $\tau_{21}$ cannot be simultaneously estimated by fitting alone. Fluorescence life-time ($\tau_{20}$) was measured by SpexFluoroLog-3 spectrometer (Jobin-Yvon Inc.) equipped with a Hamamatsu R928 photomultiplier tube and double excitation and emission mono-chromator with wavelength resolution of $\pm 1$ nm. The large values of $\tau_{10}$ and $\tau_{31}$ of H$_2$TPP with respect to H$_2$TPP(OH)$_4$ doped BAG makes it to saturate at lower intensities.

Table 1: Spectroscopic data of H$_2$TPP and H$_2$TPP(OH)$_4$ doped in BAG: Molar concentration ($\gamma$), Sample thickness (L), Phosphorescence life time ($\tau_{10}$), Fluorescence life time ($\tau_{20}$), Fluorescence life time ($\tau_{31}$), Inter-system decay life time ($\tau_{21}$), Ground state absorption cross-section ($\sigma_{02}$), Excited state absorption cross-section ($\sigma_{13}$), First saturation intensity (I$_{S01}$), Second saturation intensity (I$_{S13}$).

| Physical Parameters | H$_2$TPP doped BAG | H$_2$TPP(OH)$_4$ doped BAG |
|---|---|---|
| $\gamma$ | $4\times10^{-5}$ M | $4\times10^{-5}$ M |
| $\alpha_0$L | 0.6 | 1.6 |
| L | 88$\pm$4 μm | 67$\pm$5 μm |
| $\tau_{10}$ | $5.1\times10^{-4}$ s | $1.5\times10^{-5}$ s |
| $\tau_{20}$ | $2.05\times10^{-9}$ s | $1\times10^{-9}$ s |
| $\tau_{31}$ | $3\times10^{-7}$ s | $1\times10^{-9}$ s |
| $\tau_{21}$ | $1.1\times10^{-11}$ s | $1\times10^{-11}$ s |
| $\sigma_{02}$ | $4\times10^{-18}$ cm$^2$ | $5.2\times10^{-18}$ cm$^2$ |
| $\sigma_{13}$ | $1.7\times10^{-19}$ cm$^2$ | $7.2\times10^{-17}$ cm$^2$ |
| I$_{S01}$ | $1.48\times10^2$ W/cm$^2$ | $3.8\times10^3$ W/cm$^2$ |
| I$_{S13}$ | $5.8\times10^6$ W/cm$^2$ | $4.11\times10^8$ W/cm$^2$ |

## 4. Femto-second laser excitation

### 4.1. Model

We have estimated, the singlet to triplet decay time ($\tau_{21}$) to about 10 pico-seconds, which is larger than the incident femto-second pulse width of 130fs. Since the excitation pulse is greater in magnitude as compared to the decay time we don't expect the contribution from the triplet states. The phosphorescence life-time from T$_1$ to S$_1$ is less than the 1milli-second which is smaller than the pulse repetition time. Thus there will not be any contribution from the triplet level population and only two singlet levels will participate in the nonlinear absorption/ transmission process. In the presence of 130 femto-second laser pulses excitation at 671nm wavelength with 1 kHz repetition rate, both H$_2$TPP and H$_2$TPP(OH)$_4$ doped BAG act as two level systems (figure 7).

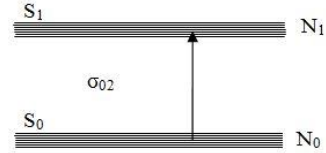

Figure 7 Energy levels of dye molecule in femto-second laser illumination.

Laser pulses with their pulse-width smaller than the excited state life-time produce a time dependent population during the absorption process [31]. The population is now a function of absorption cross-section, life-time as well as pulse-width. The excited states life-times due to spontaneous emission are found to be in the time-scale ranging from nano- to pico-seconds. Thus under femto-second laser pulse excitation, the population dynamics become independent of lifetime. The incident intensity of the light from such a source is given by equation 2.

$$I(r,t) = I_p \exp(-2r^2/\omega^2)\exp(-4\ln 2\, t^2/\tau_{FWHM}^2) \quad (2)$$

where, $I_p$ is peak intensity, $\omega$ is the spot size and $\tau_{FWHM}$ is the full width at half maximum of laser pulse. Populations decay from excited states are not considered as the excitation are much faster than the decay processes in formulating the rate equations 3.

$$n_0 + n_2 = 1 \quad (3a)$$

$$\frac{dn_0}{dt} = -\frac{\sigma_{02}I(t)}{h\nu}(n_0 - n_2) \quad (3b)$$

$$\frac{dn_2}{dt} = \frac{\sigma_{02}I(t)}{h\nu}(n_0 - n_2) \quad (3c)$$

The fractional population in the ground and excited states are obtained by integrating over pulse-widths

$$n_0 = \frac{1}{2}\left[1 + \exp\left(-2F_S^{-1}\int_{-\infty}^{+\infty}I(t)dt\right)\right] \quad (4a)$$

$$n_2 = \frac{1}{2}\left[1 - \exp\left(-2F_S^{-1}\int_{-\infty}^{+\infty}I(t)dt\right)\right] \quad (4b)$$

Here, $F_S = h\nu/\sigma_{01}$ is the saturation fluence. The absorption coefficient in terms of absorption cross-section is $\alpha = N(n_0 - n_2)\sigma_{02}$. The incident pulse energy on the sample is given by

$$E(0) = I_p(0)\frac{\pi\omega^2}{2}\sqrt{\frac{\pi}{4\ln 2}}\tau_{FWHM} \quad (5)$$

Transmitted energy from the sample is

$$E(L) = I_p(0)\frac{\pi\omega^2}{2}\sqrt{\frac{\pi}{4\ln 2}}\tau_{FWHM}\exp(-\alpha_1 L) \quad (6)$$

The overlapping of excitation mode and absorption mode is shown in the Figure 8. While H$_2$TPP doped BAG absorption peak is to the left side of the excitation pulse peak, the peak of the absorption profile of H$_2$TPP(OH)$_4$ doped BAG is to its right side due to the presence of (OH)$^-$ group. In both cases, excitation pulse are well within the absorption band

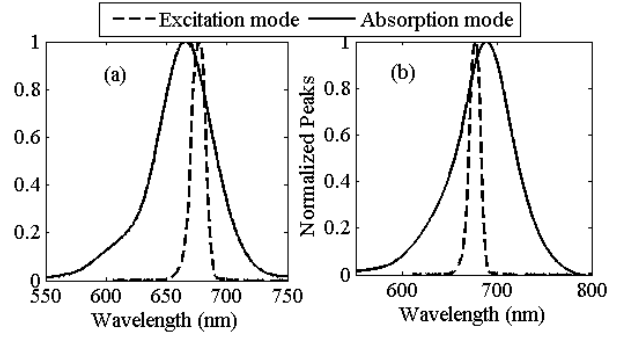

Figure 8 Central wavelength of laser pulses chosen so as to have a maximum overlap of excitation pulse profile with that of absorption profile of: (a) H$_2$TPP (4x10$^{-5}$M) doped BAG and (b) H$_2$TPP(OH)$_4$ (4x10$^{-5}$M) doped BAG.

Experimental setup used for SBT under femto-second laser excitation is shown in Figure 9. We allowed only pulses centered around 671nm from the OPA TOPAS to be incident on the sample and the rest were blocked using laser filters. To increase the range of incident intensities for SBT, we have used lens. Neutral density filters are used as intensity attenuators to vary the illumination intensity at the sample. We took care to avoid spurious reflections from the optical elements by inserting irises at appropriate position thereby improving signal to noise ratio. The fluctuation in incident intensity is monitored by reference diode (PD$_2$).

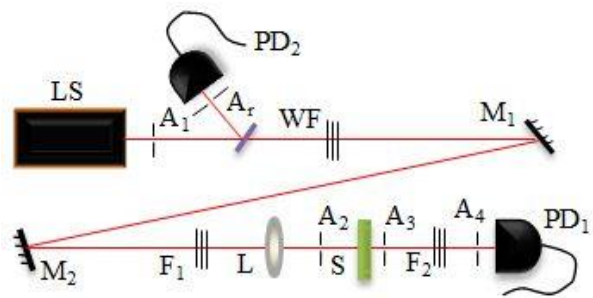

Figure 9 Experimental setup of SBT: LS → Laser source, WF → Wavelength filters, F → Neutral density filters, M$_i$ → Mirror, , L$_i$ → Lens, S → Sample, PD$_i$ → Photo diode, A$_i$ → Iris.

Figure 10 is a plot of SBT transmittance curve of H$_2$TPP (4x10$^{-5}$M) doped BAG and corresponding population redistribution is given in figure 11. The absorption cross-section is estimated to be $\sigma_{02}=1.5\times10^{-18}$ cm$^2$ which is different from the estimated value of $2.5\times10^{-18}$ cm$^2$ (Table 1) due to CW laser excitation. It is expected because of larger line-width of the probing femto-second laser as compared to cw laser beam centred on 671nm. We have seen bleaching of the sample at fluence value of 170 mJ/cm$^2$.

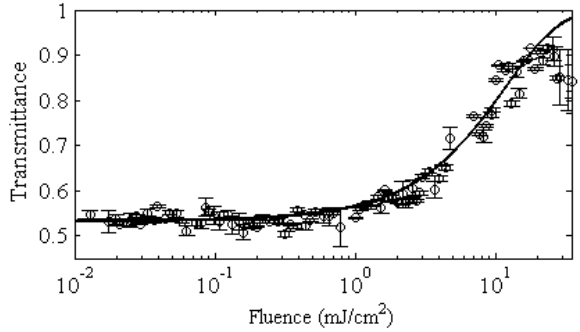

Figure 10 SBT of H$_2$TPP (4x10$^{-5}$M) doped BAG in presence of femto-second laser excitation.

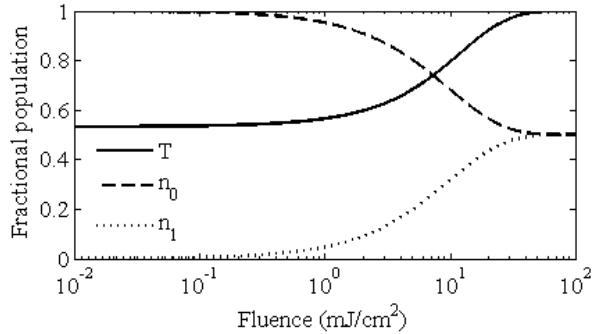

Figure 11 Fractional population distribution of H$_2$TPP (4x10$^{-5}$M) doped BAG in its energy levels.

Figure 12 is a plot of SBT curve of H$_2$TPP(OH)$_4$ doped BAG and corresponding population redistribution is shown in figure 13. The estimated absorption cross section is $\sigma_{02}=8.1\times10^{-18}$ cm$^2$ which is different from the estimated value by CW laser excitation by $2.9\times10^{-18}$ cm$^2$. Large peak power of femto-second pulses also induces optical nonlinearity in the host medium [32] viz., boric acid at the fluence value of 70 mJ/cm$^2$. In figures 10 and 12 we can see the deviation of experimental data from theoretical fit at high fluence, due to contribution to optical nonlinearity from boric acid medium.

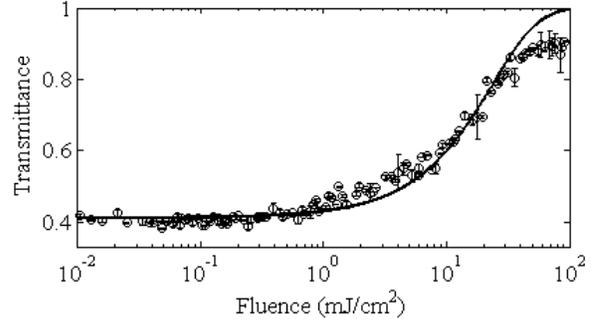

Figure 12 SBT of H$_2$TPP(OH)$_4$ (4x10$^{-5}$M) doped BAG in presence of femto-second laser excitation.

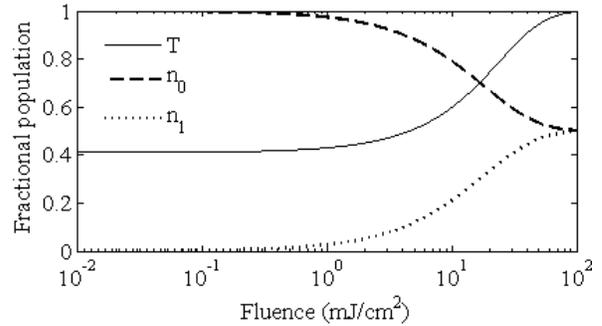

Figure 13 Fractional population distribution of H$_2$TPP(OH)$_4$ (4x10$^{-5}$M) doped BAG in femto-second laser excitation.

## 5. Conclusion

Rate equation approach under steady state condition is used to understand SBT curve. We studied the spectroscopic properties of tetra-phenyl porphyrin and its OH$^-$ group derivative prepared as thin films sandwiched between two microscope glass slides. In presence of cw laser, dye system follows the four level model. While free base tetra phenyl porphyrin molecule shows DSA behavior, the addition of OH$^-$ group changes it to RSA. Therefore H$_2$TPP can be used as saturable absorber and OH$^-$ group addition we can use it as optical limiter.

H$_2$TPP and H$_2$TPP(OH)$_4$ doped BAG act as two level system under femto-second laser pulse excitation. Analytically we have derived the time dependent fractional population by rate equation approach and then obtain the transmittance curves by fitting of the experimental data. Using SBT, we have estimated the absorption cross-sections and these values are within the one order variation with the values estimated by CW laser. This could be due to the fact that femto-second pulses have broader line-width leading to larger cross-section.

In this article, we have shown that how one can obtain absorption cross-sections and lifetimes by SBT. Through this technique we have access to larger range of incident intensity as opposed to limited range available in popular z-scan technique. The spot-size remain the same in this technique hence we would be dealing with the same set and number of molecules throughout the intensity range probed, as opposed to Zscan where the number of molecules studied changes with the intensity. Further, manipulation of cross-section and lifetimes is achieved either through the choice of host medium or introducing some functional group which modifies the microscopic parameters of the system. These kind of engineering would pave the way for better design of the materials suited for a particular application

**Acknowledgement**: Authors thanks CIF Pondicherry University for providing femto-second laser facility.